\documentclass[aps,showpacs]{revtex4}

\usepackage{amsmath}
\usepackage{amssymb}
\usepackage{graphicx}
\usepackage{dcolumn}
\usepackage{bm}
\usepackage{epstopdf}

\newcommand{\ga}{\gamma}
\newcommand{\Ga}{\Gamma}

\newcommand{\prt}{\partial}

\begin{document}

\title{
Development of instability of dark solitons generated by a flow of Bose-Einstein condensate past a
concave corner}

\author{A. M. Kamchatnov}
\email{kamch@isan.troitsk.ru}
\author{S. V. Korneev}
\email{svyatoslav.korneev@gmail.com}

\affiliation{Institute of Spectroscopy, Russian Academy of Sciences, Troitsk,
Moscow Region, 142190, Russia}

\date{\today}

\begin{abstract}
The stability of dark solitons generated
by a supersonic flow of a Bose-Einstein condensate past a concave corner (or a wedge) is studied.
It is shown that solitons in the dispersive shock wave generated at the initial moment of time
demonstrate a snake instability during their evolution to stationary curved solitons. Time of
decay of soliton to vortices agrees very well with analytical estimates of the instability
growth rate.
\end{abstract}


\maketitle

Flow of quantum liquids has attracted much attention since discovery of superfluidity.
At first it was discussed mainly in relation with the existence of the critical flow
velocity past obstacles (or boundaries of capillary tubes) above which superfluidity is
lost. According to Landau criterium it occurs at the velocity corresponding to the
threshold of generation of linear excitations in the fluid (for instance, rotons in HeII).
However, it was found that the experimental value of the critical velocity in HeII was
much less than that predicted by Landau and this contradiction between the theory and
the experiment was explained by Feynman who indicated that the threshold for generation
of nonlinear excitations, namely vortices, is much less than the Landau value and
agrees, at least semi-quantitatively, with the experimental value. Later these predictions
were confirmed by analytical and numerical study of a superfluid flow modeled by the
Gross-Pitaevskii equation for a weakly interacting Bose gas \cite{fpr-1992,WMCA}.
As was found in these studies,
a 2D flow past a disk-shaped obstacle with the radius not much greater than the healing length
looses its superfluidity at velocity equal to $v_{cr}=\sqrt{2/11}c_s\cong0.43c_s$, $c_s$ being
the sound velocity, i.e. the velocity of Bogoliubov excitations in the long wavelength limit.
Here $v_{cr}$ equals to the threshold velocity above which vortices are generated by the
flow at the boundary of the obstacle. It was noticed also in the numerical simulations that
a stationary wave pattern is generated by a supersonic flow past an obstacle and this
pattern consists of two different regions separated by the Mach cone.
These theoretical findings got new interest after experimental realization of the
flow of Bose-Einstein condensate (BEC) of dilute atomic gas past obstacles \cite{caruso}.
In theoretical works motivated by these experiments it was shown that the region outside
the Mach cone is occupied by the so called
``ship waves'' formed due to interference of Bogoliubov waves radiated by the obstacle
by means of Cherenkov effect \cite{caruso,gegk,ship2}, whereas in the region inside the Mach cone
a fan of oblique dark solitons is located \cite{ek06,egk-2006} and these oblique solitons
are effectively stable if the flow velocity is large enough \cite{kp08}.

However, if we consider flow of BEC past an extended obstacle, the wave pattern can change
drastically. In particular, the ship wave not far from the obstacle's boundary becomes
nonlinear and transforms into a stationary dispersive shock wave, i.e., in some
approximation, into a lattice of curved dark solitons \cite{slender}. In vicinity of the
boundary of the obstacle both the local sound velocity and the direction of the flow
are changed compared with their values in the incident flow and, hence, the question
of stability of dark solitons in the dispersive shocks should be reconsidered.
Their instability was noticed in numerical simulations of the highly supersonic flow
past a slender wedge \cite{slender} and it became crucially important for
moderately supersonic flows past not so slender wedges when the instability
destroyed very fast the dispersive shock pattern \cite{hi-2009}. The aim of this
paper is to give theoretical estimates of the instability growth rate
in such flows and to confirm the theory by numerical simulations.

First, we notice that if the obstacle is large and not too slender, then
transition to a stationary state formed by the ``curved solitons'' is accompanied by
propagation of bending waves along such solitons; these waves include unstable modes
which evolution can lead eventually to decay of oblique solitons to vortex-antivortex pairs.
Let at the initial moment of time $t=0$ the condensate be confined in the region outside
the trap potential having a form of a corner with one its wall located along negative $x$ axis
$(-\infty<x<0)$ and another one along the line $y=\tan\alpha\cdot x$, $\alpha$ being the
angle of the slope $(0<\alpha<\pi/2)$. The flow with velocity $M$ (equal to the Mach number
in our standard non-dimensional units with the flow velocity measured in units of the sound
velocity) along the $x$ axis is switched on at $t=0$. Evidently, in vicinity of the sloping
boundary the wave appears which is equivalent to one
generated by a piston moving with velocity $M\sin\alpha$; hence, a non-stationary dispersive shock wave
is generated \cite{slender,hi-2009,hae-08} with wave crests parallel to the sloping boundary.
If $M\sin\alpha<\sqrt{n_0}$, $n_0$ being the density of incident BEC, then the waves
closest to the boundary can be considered as oblique solitons \cite{egk-2006}.
However, these oblique solitons are cut at their edge located about the line
$y=-\cot\alpha\cdot x$.
These their edges are not attached to the obstacles and, hence, cannot become effectively
stable due to transition from their absolute instability to a convective one at large enough
flow velocity (see \cite{kp08}).
Therefore these oblique solitons must change their form transforming to ``curved solitons''
studied analytically in \cite{slender} for $M\gg1,$ $\alpha\ll1$. Hence, bending waves start their
propagation along oblique solitons and we suppose that the component of the wave packet with
maximal value of the instability growth rate $\Gamma$ leads to decay of oblique solitons
to vortices observed numerically in \cite{hi-2009}.
\begin{figure}[t]
\includegraphics[width=8cm]{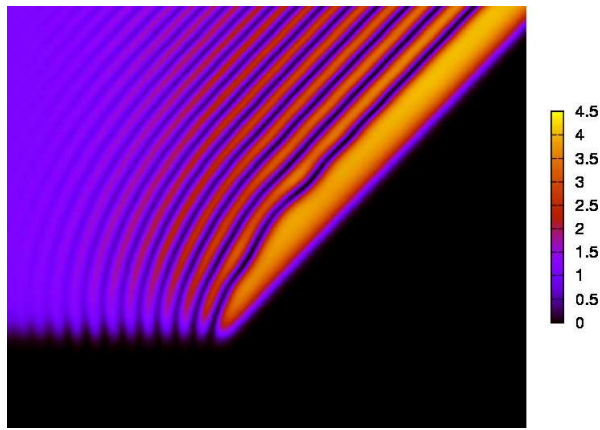}
\caption{Density plot of the wave pattern generated by the flow of BEC with the
Mach number $M=2.5$ past a corner with the angle $\alpha=52^\circ$ at the moment of
evolution $t=8.22$. The oblique solitons of the dispersive shock are modulated by
the wave of bending. These wave packets include the unstable modes which lead
eventually to decay of oblique solitons to vortices.
}
\label{fig1}
\end{figure}

To perform quantitative comparison of analytical estimates with results of numerical simulations,
we define a numerical time $T(\alpha)$ of the instability growth as follows. Direct numerical
solution of the GP equation
\begin{equation}\label{1-1}
    i\psi_t+\tfrac12(\psi_{xx}+\psi_{yy})-|\psi|^2\psi=V(x,y)\psi
\end{equation}
with the corner-like shape of the trap potential $V(x,y)$ shows that
solitons bend in the shock with time in such a way that, for instance, the
deepest one is centered around the curve $y'=Y(x',t)$, where $x'$ is
the axis along the sloping side of the corner and $y'$ is the axis normal to $x'$, that is
$(x',y')$ axes are obtained by counter-clockwise rotation of $(x,y)$ axes to the angle $\alpha$
and $Y(x',t)$ has a meaning of the distance of the soliton's trough line from the boundary
at the moment $t$.
A typical shape of this wave pattern is shown in Fig.~1 and one can see that a wave
packet propagates
along the soliton ($x'$ axis). Our calculations show that the packet's velocity is equal to
$M\cos\alpha$, that is this packet is built from non-propagating unstable modes and its motion
is caused by convection due to flow of BEC along the boundary. In Fig.~2 the
forms of wave packets corresponding to different values of $\alpha$ at such moments of time
that their amplitudes reach a certain value chosen the same one for all $\alpha$. As we see,
the forms of these wave packets are very similar to each other, too, and we choose just these moments of
time as definitions of numerical values of the instability growth time $T(\alpha)$.

To estimate the analytical growth rate values, we notice that the problem of formation of
dispersive shock wave by a piston moving with a constant velocity is equivalent to well-known
problem of evolution of a step-like pulse \cite{gk87,eggk95} and its solution was discussed in
\cite{hae-08,slender}. In particular, it was found that the condensate's density between the
corner's boundary (piston) and the soliton edge of the dispersive shock wave is equal to
\begin{equation}\label{3-1r}
    n=\left(\sqrt{n_0}+\tfrac12M\sin\alpha\right)^2,
\end{equation}
where $n_0$ is the density of BEC in the incident flow. The velocity of
the deepest soliton at the edge of the shock is equal to
\begin{equation}\label{3-2r}
    V=\sqrt{n_0}+\tfrac12M\sin\alpha.
\end{equation}
These formulae determine the parameters of the soliton which stability must be studied.
Its analytical form is well known and can be expressed as
\begin{equation}\label{9-1}
\begin{split}
    &\psi_s=\Phi_s(\zeta)e^{-i n t},\quad \Psi_s(\zeta)=k\tanh(k\zeta)-iV=
    \sqrt{n_s(\zeta)}\, e^{i\varphi_s(\zeta)},\quad
    \frac{\prt\varphi_s}{\prt\zeta}=-V\left(1-\frac{n}{n_s}\right)\\
    &k=\sqrt{n-V^2},\quad \zeta=x+Vt.
    \end{split}
\end{equation}
Then the perturbed wave function reads
\begin{equation}\label{9-2}
    \psi=(\Phi_s+\phi)e^{-i n t}=(\Phi_s+\phi'+i\phi^{\prime\prime})e^{-i n t}=
    \sqrt{n_s+\delta n}\,e^{i(\varphi_s+\delta\varphi)-i n t},
\end{equation}
where the real variables $\phi'$ and $\phi^{\prime\prime}$ are related with perturbations of the
density $\delta n$ and the phase $\delta\phi$ by the formulae
\begin{equation}\label{9-3}
    \phi'=\frac{\cos\varphi_s}{2\sqrt{n_s}}\delta n+\sqrt{n_s}\,\sin\varphi_s\cdot\delta\varphi,
    \quad
    \phi^{\prime\prime}=\frac{\sin\varphi_s}{2\sqrt{n_s}}\delta n-\sqrt{n_s}\,\cos\varphi_s\cdot\delta\varphi.
\end{equation}
Substitution of (\ref{9-2}) into (\ref{1-1}) and linearization of the resulting equation
with respect to $\phi$ gives
\begin{equation}\label{9-4}
    i\phi_t+n \phi+\tfrac12(\phi_{xx}+\phi_{yy})-(2|\Phi_s|^2\phi+\Phi_s^2\phi^*)=0,
\end{equation}
where asterisk denotes complex conjugation. We are interested in propagation of a harmonic
disturbance of the soliton along $y$ axis. As usual in linear problems, such a disturbance
can be represented in a complex form, $\delta n,\,\delta\varphi\propto\exp(\Ga t+ipy)$ or
$\phi=(\phi'+i\phi^{\prime\prime})\exp(\Ga t+ipy)$,
$\phi^*=(\phi'-i\phi^{\prime\prime})\exp(\Ga t+ipy)$, where we
keep previous notation for the amplitudes $\phi'=\phi'(\zeta),\,
\phi^{\prime\prime}=\phi^{\prime\prime}(\zeta)$.
For this form of the disturbance, Eq.~(\ref{9-4}) reduces to
\begin{equation}\label{9-5}
    i\Ga\phi+iV\phi_{\zeta}+n \phi+\tfrac12\phi_{\zeta\zeta}-\tfrac12 p^2\phi-
    (2|\Phi_s|^2\phi+\Phi_s^2\phi^*)=0.
\end{equation}
Substitution of (\ref{9-1}) and separation of real and imaginary parts yields the system
\begin{equation}\label{10-1}
    \left(
      \begin{array}{cc}
        L_{11} & L_{12} \\
        L_{21} & L_{22} \\
      \end{array}
    \right)
    \left(
      \begin{array}{c}
        \phi' \\
        \phi^{\prime\prime} \\
      \end{array}
    \right)= \Ga \left(
                   \begin{array}{c}
                     \phi' \\
                     \phi^{\prime\prime} \\
                   \end{array}
                 \right)
\end{equation}
where
\begin{equation}\label{10-2}
    \begin{split}
    L_{11}&=-V\prt_{\zeta}-2kV\tanh(k\zeta),\quad
    L_{12}=-\tfrac12\prt_{\zeta\zeta}-n +n_s+2V^2+\tfrac12p^2,\\
    L_{21}&=\tfrac12\prt_{\zeta\zeta}+n-3n_s+2V^2-\tfrac12p^2,\quad
    L_{22}=-V\prt_{\zeta}+2kV\tanh(k\zeta).
    \end{split}
\end{equation}
This eigenvalue problem was studied in \cite{kt-1988} and we used it for numerical
calculation of the instability growth rate $\Gamma(p)$
with the background density $n$ and the soliton's velocity  $V$ given by (\ref{3-1r})
and (\ref{3-2r}), respectively, for $M=2.5$ and several values of the angle
$\alpha$. These values of $\alpha$ as well as corresponding values of the background density and
solitons velocities are given in the first three columns of Table 1. The instability
growth rate depends on the wave number $p$ and its maximal value $\Gamma_{max}$ for each
$\alpha$ is indicated in the fourth column of Table 1. As we see, it decreases with growth of
$\alpha$; for convenience we have normalized its values dividing them by the value at
$\alpha=52^{\circ}$:
\begin{equation}\label{3-3r}
    \ga_{theor}(\alpha)=\frac{\Gamma_{max}(\alpha)}{\Gamma_{max}(52^{\circ})}.
\end{equation}
In the last two columns we compare these theoretical predictions with the corresponding
numerical characteristics defined as
\begin{equation}\label{4-1r}
    \ga_{num}(\alpha)=\frac{T(52^{\circ})}{T(\alpha)}
\end{equation}
where $T(\alpha)$ is the defined above instability growth time. One can see very good
agreement which confirms our approach to the problem of instability of oblique solitons.

\begin{figure}[t]
\includegraphics[width=8cm]{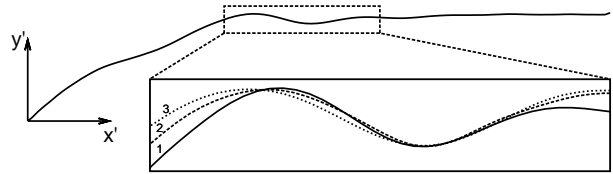}
\caption{Line of trough of the deepest dark soliton generated by the flow of BEC with
$M=2.5$ at the moment of time $t=8.22$ past a corner with the angle $\alpha=52^\circ$.
In the insert the wave packets with the same amplitude are shown for the angles
(1) $\alpha=52^\circ$, (2) $\alpha=37^\circ$, (3) $\alpha=32^\circ$. The corresponding
moments of time are chosen as times $T(\alpha)$ of the instability growth.
}
\label{fig2}
\end{figure}

\smallskip
\begin{center}
Table I. Analytical and numerical estimates of instability growth rate
of oblique solitons (Mach number $M=2.5$).
\smallskip
\begin{tabular}{|c|c|c|c|c|c|c|c|}
\hline
$\alpha$  & $n$ & $V$ & $\Gamma_{max}$ & $T(\alpha)$ & $\ga_{num}(\alpha)$ & $\ga_{theor}(\alpha)$\\
\hline
$52^{\circ}$ & 4.16 & 0.07 & 1.09 & 8.22 & 1 & 1 \\
\hline
$47^{\circ}$ & 3.82 & 0.13 & 0.99 & 9.1 & 1.09 & 1.11 \\
\hline
$42^{\circ}$ & 3.48 & 0.19 & 0.89 & 10.34 & 1.22 & 1.26 \\
\hline
$37^{\circ}$ & 3.12 & 0.26 & 0.79 & 11.41 & 1.38 & 1.39 \\
\hline
$32^{\circ}$ & 2.80 & 0.35 & 0.68 & 13.03 & 1.60 & 1.59 \\                                                                                         \hline
\end{tabular}
\end{center}
\smallskip

Thus, our results show that the size of the obstacle is crucially important for realization
of transition to effectively stable dark soliton states. In numerical experiment of
Ref.~\cite{egk-2006} and theoretical estimates of Ref.~\cite{kp08} the size of the
obstacle did not play any essential role. However, if a characteristic size $l$ of
the obstacle is such that the time $l/v_{conv}$ (where $v_{conv}$ is the velocity of convection)
is greater than the inverse instability growth rate $\Gamma_{max}^{-1}$, then the
unstable wave packet has enough time for evolution into its nonlinear stage with
formation of vortex-antivortex pairs. Hence, if we wish to observe effectively stable
dark solitons in the flow of BEC past a concave corner or a wedge, then the parameters
of the flow must satisfy the condition
\begin{equation}\label{5-1}
    \frac{l}{M\cos\alpha}\ll\frac1{\Gamma_{max}}.
\end{equation}
This condition was fulfilled in most of numerical experiments of Ref.~\cite{slender} but was not
fulfilled in numerical experiments of Ref.~\cite{hi-2009} what has lead to fast decay of
oblique solitons into vortices. Our estimates show that the location of the ``seed''
disturbance where the first vortices are formed is determined by the distance
$M\cos\alpha/\Gamma_{max}$ along the inclined wall of the corner.

It is worth noticing also that importance of existence of such a period in evolution
of dark solitons that the unstable modes have enough time for their development into
vortices manifested itself also in evolution of dark ring solitons \cite{ring-1,ring-2}:
the instability develops effectively near the turning point of the ring soliton
evolution when its parameters do not practically change and the mode with maximal corresponding
value of $\Gamma$ dominates in the evolution of the soliton's shape.x

We thank Yu.G.Gladush for useful discussions.
This work was supported by RFBR (grant 09-02-00499).

\end{document}